\documentclass[universe,article,accept,pdftex,moreauthors]{Definitions/mdpi}

\firstpage{1}
\pubvolume{1}
\issuenum{1}
\articlenumber{0}
\pubyear{2024}
\copyrightyear{2024}
\datereceived{3 April 2024}
\daterevised{7 May 2024} 
\dateaccepted{13 May 2024}
\datepublished{  }
\hreflink{\linebreak} 
\pdfoutput=1 

\usepackage{bm} 

\Title{The Rotation of Classical Bulges in Barred Galaxies in the Presence of Gas}

\TitleCitation{The Rotation of Classical Bulges in Barred Galaxies in the Presence of Gas}

\Author{Rubens E. G. Machado$^{1}$\orcidA{}*, Kenzo R. Sakamoto $^{1}$\orcidB{}, Andressa Wille$^{1}$\orcidC{} and Gustavo F. Gonçalves$^{1}$\orcidD{}}

\AuthorNames{Rubens E. G. Machado, Kenzo R. Sakamoto, Andressa Wille and Gustavo F. Gonçalves}

\AuthorCitation{Machado, R. E. G.; Sakamoto, K. R.; Wille, A.; Gonçalves, G. F.}

\address[1]{%
$^{1}$ \quad Departamento Acad\^emico de F\'isica, Universidade Tecnol\'ogica Federal do Paran\'a, Av. Sete de Setembro 3165, Curitiba, Brazil\\}

\corres{Correspondence: rubensmachado@utfpr.edu.br}

\abstract{Barred galaxies often develop a box/peanut pseudobulge, but they can also host a nearly spherical classical bulge, which is known to gain rotation due to the bar. We aim to explore how the presence of gas impacts the rotation of classical bulges. We carried out a comprehensive set of hydrodynamical $N$-body simulations with different combinations of bulge masses and gas fractions. In these models, both massive bulges and high gas content tend to inhibit the formation of strong bars. For low-mass bulges, the resulting bar is stronger in cases of low gas content. In the stronger bar models, bulges acquire more angular momentum and thus display considerable rotational velocity. Such bulges also develop anisotropic velocity dispersions and become triaxial in shape. We found that the rotation of the bulge becomes less pronounced as the gas fraction is increased from 0 to 30\%. These results indicate that the gas content has a significant effect on the dynamics of the classical bulge, because it influences bar strength. Particularly in the case of the low-mass bulges (10\% bulge mass fraction), all of the measured rotational and structural properties of the classical bulge depend strongly and systematically on the gas content of the galaxy.}


\begin{document}

\section{Introduction}

In the central region of disk galaxies, a protuberance in the light distribution has been historically called a bulge. Bulges have long been understood to be similar to elliptical galaxies, since they are generally smooth spheroidal stellar structures dominated by random motion. To a first approximation, bulges even share with elliptical galaxies roughly the same $R^{1/4}$ radial profile of surface brightness. However, recent developments in the field indicate that bulges are more complicated and more diverse (see \cite{Kormendy2004} for a review) than this early picture would have suggested. Moreover, bulges in general may hold important information to understand the formation and evolution of galaxies, both regarding early phases of hierarchical mergers and also more recent but long-term dynamical processes during secular evolution.

In particular, new classes of bulges have been identified and tentatively classified due to their peculiar structural and kinematic properties, and also stellar populations. Some bulges exhibit velocity structures that are more reminiscent of disks than ellipticals \cite[e.g.][]{Gadotti2009}, in the sense that they are dominated more by rotation than by random motions. These classes of objects are sometimes called `pseudobulges' in order to distinguish them from the `classical bulges'.

Apart from disk-like bulges, a particularly interesting type of pseudobulges are the structures sometimes referred to as box/peanut (BP) bulges or box/peanut/X-shaped (BPX) bulges. Such X-shapes can be detected in observations of edge-on galaxies \cite[e.g][]{Ciambur2016, Savchenko2017}. They are now understood to be the product of bar formation and evolution. Specifically, simulations \cite{Athanassoula2005} reveal that a BPX morphology emerges when a strongly barred galaxy is viewed side-on, i.e.~when the galactic disk is seen edge-on but additionally the longer axis of the stellar bar is perpendicular to the line of sight. In $N$-body simulations, strong bars develop the peculiar peanut or X-shape when seen side-on after the buckling instability \cite{MartinezValpuesta2006, Lokas2019}. Recent analysis of cosmological simulations \cite{Anderson2024} also indicates that BP are found in galaxies with long and strong bars, and that have undergone the phenomenon of buckling in the past, regardless of galactic mass.

Classical bulges may originate as a result of major or minor mergers \cite{Hopkins2009, Hopkins2010} in the early formation history of the galaxy. On the other hand, pseudobulges may be formed later by dynamical processes connected to the disk and the bar \cite{Athanassoula2005}. The processes that lead to disk growth in general, and to bar formation in particular, may take place normally in a galaxy that already hosts a pre-existing classical bulge formed from the earlier merger history. Therefore, it is conceivable that classical and BP bulges can coexist. {In fact, barred galaxies are frequent in the local Universe \cite{MenendezDelmestre2007, Sheth2008}. Therefore, it is plausible that the coexistence of classical bulges and bars may be a common configuration, rather than a rare one.} Observationally, the decomposition \cite{Gadotti2009} of these superposed classical and BP components is more difficult than in simulations.

Regardless of the vertical structure of BP bulges, the presence of a strong stellar bar may by itself impact the evolution of a previously formed classical bulge. The phenomenon of the spin-up of classical bulges has been identified in high-resolution collisionless simulations of barred galaxies \cite{Saha2012, Saha2016}. The physical mechanism responsible for the acquisition of rotation by the bulge is the transfer of angular momentum that takes place at the resonances. Using orbital spectral analysis, \cite{Saha2012} showed that the transfer of the angular momentum to the bulge takes place mainly through resonances. These transfers of angular momentum importantly alter the orbital structure of the classical bulge. They found that Inner Lindblad Resonance (ILR) is responsible for an important part of the angular momentum exchange. However, the Outer Lindblad Resonance (OLR) and the corotation itself did not play a significant role in this context, presumably due to the fact that the bulge is much smaller than the halo.

{The role of angular momentum transfer in galactic dynamics has been recognized since the foundational work of \cite{LyndenBell1972}. The exchange of angular momentum between the bar and spheroidal components was proposed in \cite{Kormendy1979} and this exchange was measured in $N$-body simulations in \cite{Sellwood1980}. Subsequently, the detailed interplay between bar evolution and angular momentum transfer has been understood in terms of angular momentum transfer via resonant orbits \cite[e.g.][]{Hernquist1992, Athanassoula2002, Athanassoula2003, Weinberg2007} mainly with the aid of $N$-body simulations.
Resonant orbits are important because linear theory \cite{LyndenBell1972} shows that there is a direct relation between periodic perturbations to the potential and the change in angular momentum. The most relevant resonance in this context is usually the corotation resonance (CR), where the angular frequency $\Omega$ of the stars is equal to the pattern speed of the bar $\Omega_{\rm b}$. The Inner Lindblad Resonance (ILR) occurs at the radius in which the frequencies are related by $\Omega - \kappa/2 = \Omega_{\rm b}$, where $\kappa$ is the epicycle frequency. The ILR may correspond to one radius, or to two radii (inner ILR and outer ILR), or it may be undefined, depending on the inner mass distribution of the galaxy. The Outer Lindblad Resonance (OLR) corresponds to $\Omega + \kappa/2 = \Omega_{\rm b}$. Most of the angular momentum transfer within a barred galaxy potential takes place at or near these radii.
One of the essential mechanisms is that stars that were in nearly circular orbits outside the bar may become trapped into elongated orbits. In this way, angular momentum of the outer disk is diminished, because some stars in large circular orbits are captured. At the same time, the bar becomes longer and stronger for having trapped such stars. Furthermore, the orbits of stars trapped by the bar may become progressively more elongated with time. This also corresponds to a decrease in angular momentum. Finally, the bar as a whole will tend to diminish its pattern speed, again losing angular momentum.
These processes are connected to each other and to the ability of other components to absorb angular momentum. In particular, the spheroidal components (halo and bulge) tend to be more efficient at receiving angular momentum if they are dynamically colder \cite{Athanassoula2003}.
Among the spheroidal components, the dark matter halo is by far the dominant sink of angular momentum, because of its greater mass. However, the bulge is also found to absorbs a non-negligible amount of angular momentum \cite{Saha2012}.}

At the same time, the inclusion of gas in hydrodynamical simulations of barred galaxies has revealed that bar formation tends to be inhibited, or at least delayed, in the presence of gas \cite{Athanassoula2013}; furthermore, the properties of the bars depend strongly on gas fraction; bars formed in gas-rich simulated galaxies are much weaker.

In view of this, it is relevant to investigate how the addition of a gaseous component would impact the spin-up of classical bulges in barred galaxies. Specifically, we are interested in the connection between bars and bulges, possibly mediated by gas. The inclusion of gas in simulations of galactic dynamics often reveals specific phenomena that could hardly have been foreseen otherwise. Thus such simulation approach is valuable not only to clarify known open questions and to elucidate physical mechanisms, but also to uncover potentially new issues.

In this work, we explore a comprehensive set of models built to evaluate the effects of both the presence of gas and the presence of classical bulges. To achieve this, we compare models of galaxies with several combinations of these parameters. This allows us to separate the impact of gas content for each given bulge mass and, conversely, the impact of bulge mass for each given gas content. To serve as references, gasless galaxies are also simulated, as well as bulgeless galaxies. The output of these simulations is used to quantify the rotational and structural properties of the classical bulge, to study its secular evolution, and to look for patterns between the bulge features and the gas content of the galaxy.

This paper is structured as follows. In Section~\ref{sec:methods}, we describe the parameters of the initial conditions and the properties of the set of simulated galaxies. In Section~\ref{sec:results}, we present our results, in which we analyze bar strength, angular momentum transfer, the kinematics and the shape of the bulge. A discussion is provided in Section~\ref{sec:discussion} and finally we summarize our conclusions in Section~\ref{sec:conclusions}.

\section{Simulation setup}
\label{sec:methods}

In this section, we describe the creation of the initial conditions of the simulations. Each galaxy is, in principle, represented by 4 components: the dark matter halo, the stellar disk, the gas disk, and the stellar bulge. As will be discussed, some galaxies will be bulgeless, and some galaxies will be gasless.

The halo follows a Hernquist \cite{Hernquist1990} density profile:
\begin{equation} \label{eq:hern}
\rho(r) = \frac{M_{\rm h}}{2\pi} \frac{a_{\rm h}}{r} \frac{1}{(r+a_{\rm h})^3},
\end{equation}
where $M_{\rm h}=9 \times 10^{11}\,{\rm M}_{\odot}$ is the mass of the halo and $a_{\rm h}=30$\,kpc is the scale length of the halo. The halo also has a spin parameter of $\lambda=0.033$, which is consistent with typical values from cosmological simulations \cite[e.g.][]{Hetznecker2006, Rodriguez2022}. In our simulations, the halo parameters are kept the same for all models.

The stellar disk density profile is:
\begin{equation}
\rho(R,z) = \frac{M_{\rm d}}{4 \pi R_{\rm d}^2 z_{\rm d}} ~ e^{-R/R_{\rm d}} ~{\rm sech}^2{\left(\frac{z}{z_{\rm d}}\right)},  
\end{equation}
where $M_{\rm d}$ is the mass of the disk, $R_{\rm d}=2.5$\,kpc is the radial scale length and $z_{\rm d}=0.5$\,kpc is the vertical scale length. The gas follows the same profile as the stellar disk.

The bulge is initially spherical, with the same Hernquist density profile of equation~\ref{eq:hern}, but with a scale length of 0.5\,kpc. This translates into a half-mass radius of 1.2\,kpc for the bulge. The bulge masses are explained next.

\begin{table}[!h]
\caption{Parameters of the initial conditions.}
\label{tab1}
\newcolumntype{C}{>{\centering\arraybackslash}X}
\begin{tabularx}{\textwidth}{CCCCCCC}
\toprule 
\textbf{Label} & \textbf{Bulge fraction} & \textbf{Gas fraction} & $\bm{M_{\rm h}}$ & $\bm{M_{\rm d}}$ & $\bm{M_{\rm b}}$ & $\bm{M_{\rm g}}$ \\
 ~ & (\%) & (\%) & ($10^{10}\,{\rm M}_{\odot}$) & ($10^{10}\,{\rm M}_{\odot}$) & ($10^{10}\,{\rm M}_{\odot}$) & ($10^{10}\,{\rm M}_{\odot}$)  \\
\midrule
B0G0 & 0  & 0  & 90 & 5.2 & 0   & 0    \\
B0G1 & 0  & 10 & 90 & 4.7 & 0   & 0.5  \\
B0G2 & 0  & 20 & 90 & 4.2 & 0   & 1    \\
B0G3 & 0  & 30 & 90 & 3.6 & 0   & 1.6  \\[0.5em]
B1G0 & 10 & 0  & 90 & 4.2 & 0.5 & 0    \\
B1G1 & 10 & 10 & 90 & 4.2 & 0.5 & 0.5  \\
B1G2 & 10 & 20 & 90 & 3.7 & 0.5 & 1    \\
B1G3 & 10 & 30 & 90 & 3.3 & 0.5 & 1.4  \\[0.5em]
B2G0 & 20 & 0  & 90 & 4.2 & 1   & 0    \\
B2G1 & 20 & 10 & 90 & 3.8 & 1   & 0.4  \\
B2G2 & 20 & 20 & 90 & 3.3 & 1   & 0.9  \\
B2G3 & 20 & 30 & 90 & 2.9 & 1   & 1.3  \\[0.5em]
B3G0 & 30 & 0  & 90 & 3.7 & 1.5  & 0   \\
B3G1 & 30 & 10 & 90 & 3.3 & 1.5  & 0.4 \\
B3G2 & 30 & 20 & 90 & 2.9 & 1.5  & 0.8 \\
B3G3 & 30 & 30 & 90 & 2.6 & 1.5  & 1.1 \\[0.5em]
\bottomrule
\end{tabularx}
\end{table}

\begin{figure}[!h]
\includegraphics[]{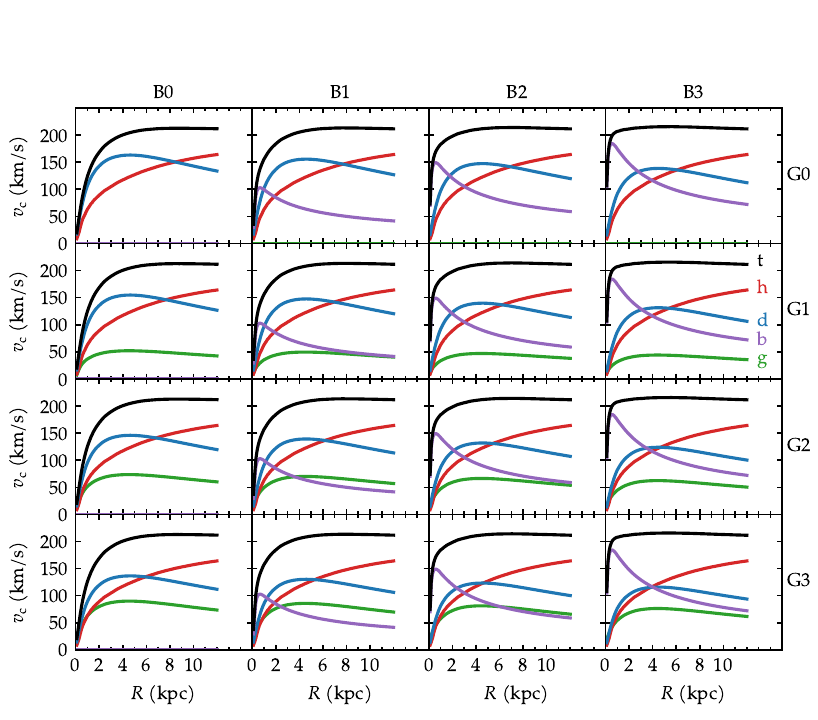}
\caption{Circular velocity curves of the initial conditions for the components: total (black), halo (red), stellar disk (blue), bulge (purple), and gas (green). The columns correspond to different bulge fractions; the rows correspond to different gas fractions.}
\label{fig01}
\end{figure}

In order to study the combined effects of the bulge and the gas, we need to create initial conditions with different combinations of masses for these components. However, to allow for fair comparisons between models, the total mass of the galaxy should be the same in all cases. Therefore, we adopted a fixed total mass of $M_{\rm baryons} = 5.2 \times 10^{10}\,{\rm M}_{\odot}$ for the baryons, i.e.~for the sum of masses in the gas, stellar disk, and bulge. The mass in the disk component may be in the form of stars or gas. In models that contain gas, a gas fraction was defined as:
\begin{equation}
f_{\rm gas} = \frac{M_{\rm gas}}{M_{\rm disk}+M_{\rm gas}},
\end{equation}
where $M_{\rm disk}$ is the mass of the stellar disk. The adopted range of gas fraction was from 0 to 30\%. Likewise, part of the total baryonic mass may be in the form of a bulge. The bulge mass fraction was established as:
\begin{equation}
f_{\rm bulge} = \frac{M_{\rm bulge}}{M_{\rm disk}+M_{\rm gas}+M_{\rm bulge}},
\end{equation}
The adopted range of bulge mass fraction was also from 0 to 30\%. For each given $f_{\rm bulge}$ of 0, 10, 20 and 30\%, we created galaxies with given $f_{\rm gas}$ of 0, 10, 20 and 30\%. The nomenclature for all the 16 combinations of initial conditions is presented in Table~\ref{tab1}, {along with the values of the masses for all components}. For example, galaxy B0G0 is bulgeless and gasless, while galaxy B1G2 has 10\% bulge and 20\% gas. We sometimes refer to `the set of B1 models', meaning all the 4 gas fractions (G0, G1, G2 and G3) within that set. Figure~\ref{fig01} displays the circular velocity curves of the initial conditions.

The initial conditions were realized with the methods from \cite{Springel2005b}. The number of particles in the halo is always $N_{\rm h}=10^6$. The number of particles in the gas is always $N_{\rm g}=2\times10^5$. The number of particles in the stellar disk and in the bulge are always $N_{\rm d+b}=2\times10^5$, distributed proportionately. {A comparison regarding mass resolutions is presented in Appendix~\ref{app:resolution}.} The simulations were carried out with the Gadget-2 code \cite{Springel2005} for 10\,Gyr, with a gravitational softening length of 0.05\,kpc. Star formation was not used in these simulations, meaning that the gas fraction of a given model remains constant throughout the evolution.

\section{Results}
\label{sec:results}

\subsection{Bar evolution}
\label{sec:bar}

In this section, we analyze the evolution of the bar strength. The bar is the driving mechanism behind the spin-up of the classical bulges. Thus it is necessary to characterize bar formation and evolution in all models, to establish how it connects to bulge mass and gas mass.

\begin{figure}
\includegraphics[]{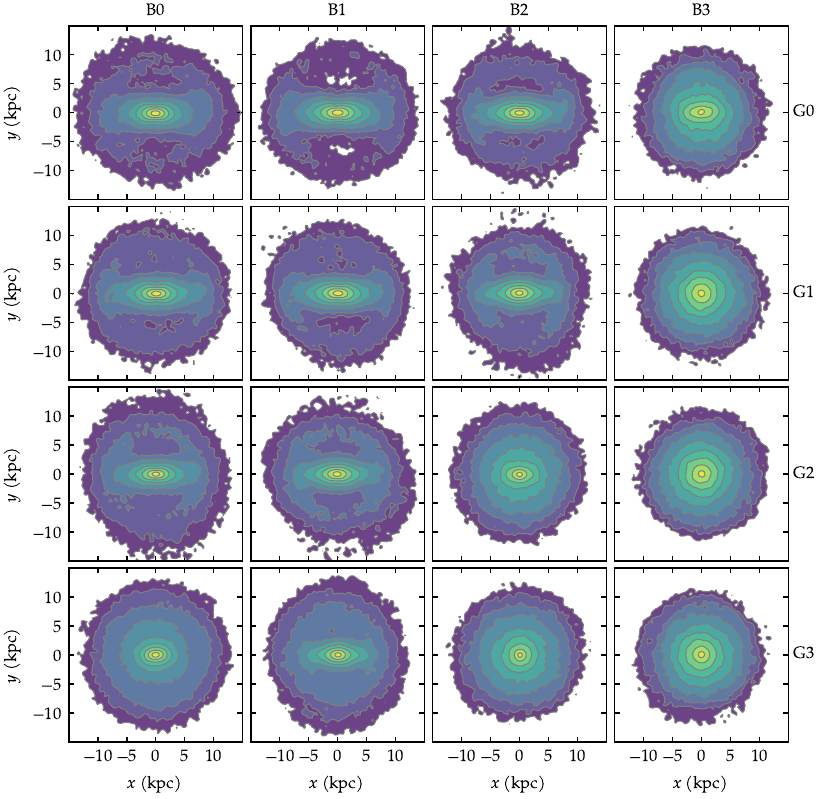}
\caption{Face-on projections of the stellar disk mass at $t=10$\,Gyr. The columns correspond to different bulge fractions; the rows correspond to different gas fractions.}
\label{fig02}
\end{figure}

Some galaxies develop strong bars, while others do not. The overall morphology of the stellar disk at the end of the simulation ($t=10$\,Gyr) can be seen in Figure~\ref{fig02}, which exhibits the face-on projection of the mass distribution. The columns of Figure~\ref{fig02} display the different bulge fractions from 0 to 30\%, while the rows display the different gas fractions, also from 0 to 30\%. Some galaxies remain nearly as axisymmetric as they were in the initial conditions. Others develop a mild oval by the end of the simulation. And others form a very strong elongated bar. It is already qualitatively clear from Figure~\ref{fig02} that the strongest bars are generally found towards the upper left frames (bulgeless, gasless), while the non-barred galaxies are found towards the lower right panels (massive bulge, high gas content). This means that both elements (gas and bulge) seem to generally inhibit bar formation.

\begin{figure}
\includegraphics[]{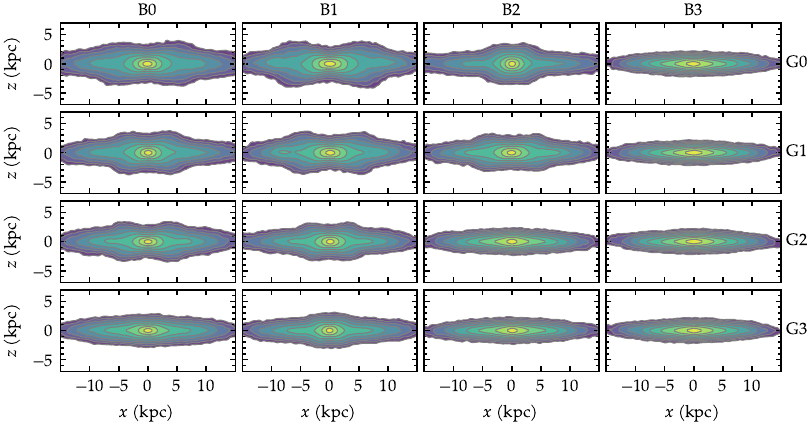}
\caption{Edge-on projections of the stellar disk mass at $t=10$\,Gyr. The columns correspond to different bulge fractions; the rows correspond to different gas fractions.}
\label{fig03}
\end{figure}

For completeness, Figure~\ref{fig03} displays the edge-on projection of the mass distribution of the stellar disks, also at $t=10$\,Gyr. The most prominent feature in this projection are the box/peanut bulges seen in the galaxies with strong bars. In Figure~\ref{fig03} (as well as in Figure~\ref{fig02}), the bars were rotated such that their longer axes are aligned with the $x$ axis, i.e.~the bars are seen side-on in Figure~\ref{fig03}. While the study of box/peanut bulges is not itself the main focus of the present paper, it is relevant to bear in mind that they are present in strongly barred galaxies and coexist spatially with the classical bulges.

\begin{figure}
\includegraphics[]{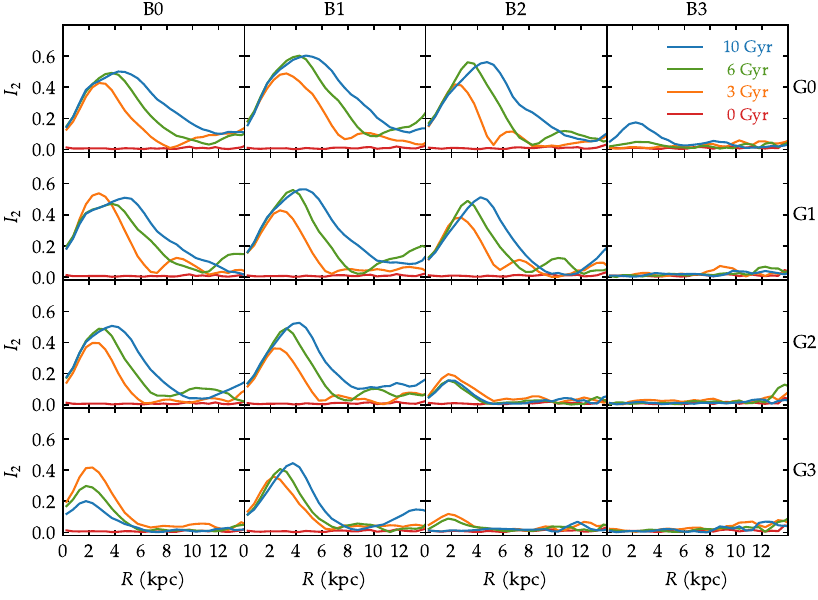}
\caption{The relative amplitude of the $m=2$ Fourier component of the stellar disk mass distribution as a function of radius for all times. The columns correspond to different bulge fractions; the rows correspond to different gas fractions.}
\label{fig04}
\end{figure}

In order to quantify the evolution of bar strength, we adopted the usual method \cite[e.g.][]{Machado2010, Athanassoula2013} of measuring the relative amplitude of the $m=2$ mode of the Fourier decomposition of the projected mass. The quantity $I_2$ is measured within each annulus of cylindrical radius $R$:
\begin{equation}
I_2 = \frac{\sqrt{a_2^2+b_2^2}}{a_0},
\end{equation}
{where $a_2$ and $b_2$ are the Fourier coefficients of the $m=2$ mode, while $a_0$ corresponds to the axisymmetric $m=0$ mode. The $a_m$ and $b_m$ coefficients are generically defined as:}
\begin{equation}
{a_m (R) = \sum _{i=0}^{N_{R}}~m_{i}~\cos (m\theta_i), ~ m=0, 1, 2, ...}
\end{equation}
\begin{equation}
{b_m (R) = \sum _{i=0}^{N_{R}}~m_{i}~\sin (m\theta_i), ~ m=1, 2, ... }
\end{equation}
{where $\theta_i$ is the azimuthal angle of each particle $i$ in the $xy$ plane, and $N_{R}$ is the number of particles within each annulus. Then we take the maximum value of the radial profiles of $I_2(R)$ to be the bar strength $A_2$ defined as:}
\begin{equation}
A_2 = \max(I_2(R)) 
\end{equation}
at each time. Figure~\ref{fig04} shows the profiles of $I_2(R)$ for all models at all times. A few selected times are highlighted with thick lines in Figure~\ref{fig04}. The intermediate curves between these times were attributed correspondingly intermediate colors. We see that bar formation takes place on a short time scale (from red to orange lines in Figure~\ref{fig04}). The radial profiles of $I_2(R)$ also indicate that the weak bars are usually also short (for instance, galaxy B3G0), which is consistent with the visual impression from Figure~\ref{fig02}. 

Regarding bulge mass, we find that strong bars develop in all bulgeless (B0) models and in the low-mass bulge cases (B1). In the case of intermediate bulge (B2), bars gradually go from strong (G0, G1) to weak (G2) to nonexistent (G3) as the gas fraction increases. Finally, for massive bulges (B3), bar formation is almost entirely suppressed, with only a weak residual elongation at the end of the gasless case B3G0.

\begin{figure}
\includegraphics[]{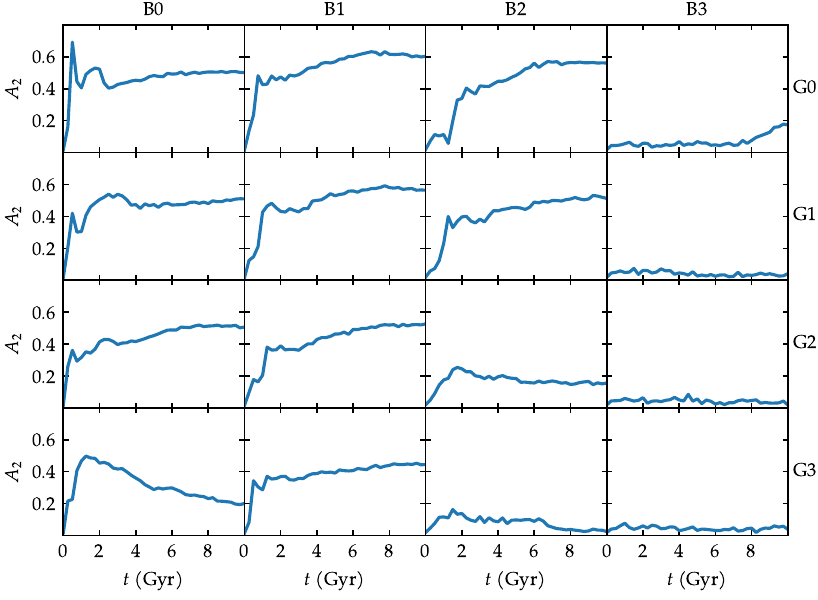}
\caption{Bar strength as a function of time. The columns correspond to different bulge fractions; the rows correspond to different gas fractions.}
\label{fig05}
\end{figure}

The time evolution of bar strength is quantified in Figure~\ref{fig05}. With few exceptions, the prevailing trend is that bars are weaker towards the right and towards the bottom of Figure~\ref{fig05}. Take, for instance, a given gas fraction, say G2: bars become weaker along the B0--B3 progression. Likewise, take a given bulge fraction, say B2: bars become weaker along the G0--G3 progression. Some exceptional cases add noise to these correlations. For example, galaxy B0G3 (bulgeless, gas-rich) is a peculiar case where the bar weakens on its own. Galaxy B3G0 (gasless, massive bulge) undergoes a mild late growth of the bar.

\begin{figure}
\includegraphics[]{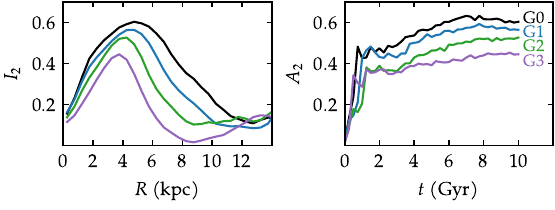}
\caption{(\textbf{a}) The relative amplitude of the $m=2$ Fourier component of the stellar disk mass distribution as a function of radius, shown for all gas fractions of the B1 set of models at $t=10$\,Gyr. (\textbf{b}) Bar strength as a function of time, shown for all gas fractions of the B1 set of models.}
\label{fig06}
\end{figure}

To inspect the effect of gas fraction more closely, we single out one specific set of models, namely B1, which has low-mass bulges. The B1 set of galaxies lends itself to clear comparisons, and thus it will be used to illustrate the main effects of gas through the paper. Figure~\ref{fig06} displays the $I_2$ of models B1 at $t=10$\,Gyr, and also the time evolution of $A_2$ for those models. In this comparison, we can clearly note the very systematic effect of varying gas fractions along the G0--G1--G2--G3 (black, blue, green, purple) progression. The increase of gas fraction by steps of 10\% systematically decreases the bar strengths.

{In order to quantify the correlations of bar strength with both gas fraction and with bulge fraction, we computed the Spearman correlation coefficients $\rho$, as well as their respective $p$-values, at $t=6$\,Gyr. The correlation between $A_2$ and $f_{\rm gas}$ was $\rho = -0.497, (p=0.050)$. The correlation between $A_2$ and $f_{\rm bulge}$ was $\rho = -0.606, (p=0.013)$. The negative coefficients naturally indicate an anti-correlation, in the sense that increasing either $f_{\rm gas}$ or $f_{\rm bulge}$ tends do decrease $A_2$. It is interesting to note that the correlation with $f_{\rm bulge}$ seems marginally more intense. This is partially due to the fact that bars were strongly suppressed in the B3 models. Consider, for instance, galaxies B3G0 and B3G1, at the top right corner of Figure~\ref{fig05}. These galaxies were prevented from having strong bars, in spite of being gasless. This brings down their values of $A_2$ at the low end of the $f_{\rm gas}$ range, contributing to the larger scatter of the $A_2$ \textit{vs.} $f_{\rm gas}$ correlation. Conversely, the G3 models did not suppress bars as fully. As a results, the $A_2$ \textit{vs.} $f_{\rm bulge}$ correlation is slightly more well-defined.}

\subsection{Angular momentum}

In this subsection, we start to quantify the rotation of the classical bulge by measuring the transfers of angular momentum.

\begin{figure}
\includegraphics[]{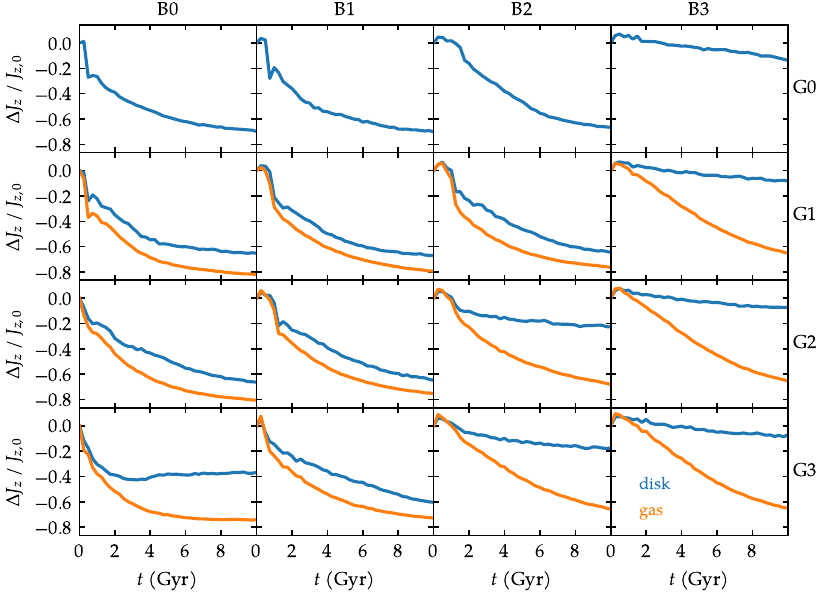}
\caption{Variation of angular momentum relative to its initial value, as a function of time, {within $R<4.2$\,kpc}, shown for the stellar disk (blue) and for the gas (orange). The columns correspond to different bulge fractions; the rows correspond to different gas fractions.}
\label{fig07}
\end{figure}

\begin{figure}
\includegraphics[]{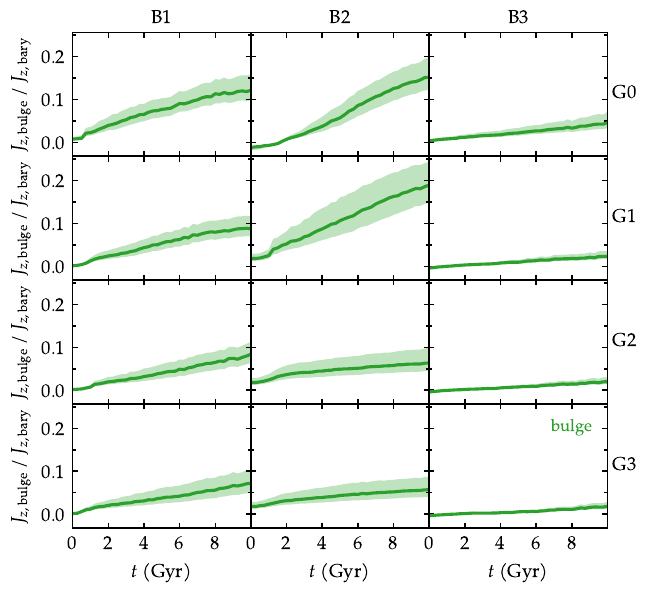}
\caption{Fractional angular momentum held by the bulge, relative to the total angular momentum of the baryons (stellar disk, gas, and bulge), as a function of time. The columns correspond to different bulge fractions; the rows correspond to different gas fractions. {Measurements were limited to $R<4.2$\,kpc, and the shaded regions correspond to $\pm20\%$ of this adopted radius.}}
\label{fig08}
\end{figure}

It is well established that bars form in galaxies as the disk loses angular momentum mainly to the halo through resonances. Angular momentum is also transferred from the inner disk to the outer disk. {(The locations of the resonances are shown in two illustrative examples in Appendix~\ref{app:resonances}.)} At first, we simply characterize the decrease of angular momentum of the disk components. Figure~\ref{fig07} displays the variation of the vertical component of the angular momentum relative to its initial value, i.e.~$(J_z - J_{z,0}) / J_{z,0}$. This is done separately for the disk (blue) and for the gas (orange) components. {In galactic disks, the outer regions tend to dominate the angular momentum. In order to highlight the relative transfers, we excluded the outer regions by restricting the measurements of angular momentum to a given radius of $R<4.2$\,kpc. This adopted radius corresponds approximately the initial half-mass radius of the exponential disks used in this paper.} The expected behavior is seen in Figure~\ref{fig07}: the most intense exchanges of angular momentum take place in the models where strong bars are formed. In the upper left panels of Figure~\ref{fig07} one finds steeper slopes, whereas in the lower right panels one finds flat slopes for the stellar disk. Thus there is a clear decrease in the values of disk angular momentum as a function of time, and this decrease is less important in the cases of massive bulges and gas-rich disks.

Regardless of the decrease in disk $J_z$, we would like to quantify the rotation of the bulge by comparing its angular momentum to the angular momentum due to all baryons (gas, stellar disk, and bulge). Thus in Figure~\ref{fig08} we measure the fractional angular momentum of the bulge relative to all baryons. Naturally, this is a small fraction, since the spatial extent of the bulge is much smaller than that of the disk, and also the bulge rotational velocities (as we will see in the next subsection) are quite smaller as well. Nevertheless, the mere fact that there is some measurable nonzero net angular momentum in the bulge indicates that it gained rotation throughout the evolution. The initial velocities of the bulge are isotropic, and thus its initial angular momentum $J_z,0$ is close to zero, apart from some very small random noise, which could be in any direction, positive or negative.

{If the entire disk were taken into account in this measument, the angular momentum of the bulge would reach at most 1--2\% of the baryonic angular momentum. By excluding the outer disk and focusing only on the inner region ($R<4.2$\,kpc), we were able to detect a clearer signal.} Figure~\ref{fig08} shows that the angular momentum of the bulge starts to grow immediately in the beginning of the simulations and continues to grow almost steadily throughout. {In the strongest cases, the angular momentum of the bulge may reach as much as 10--20\% of the angular momentum of the baryons within this $R<4.2$\,kpc region.} For a given gas fraction, the effect of the progressively larger bulge masses is not always systematic. Conversely, for one given bulge mass, the effect of the gas seems systematic. {Furthermore, to account for the apparent arbitrary choice of the radial cut of $R<4.2$\,kpc, the shaded regions of Figure~\ref{fig08} display what results would have been if the choice of radial cut had been 20\% larger or smaller than the adopted values, namely 3.4--5.0\,kpc. One finds that the dispersion is proportional to the rate of change, but the overall results are not extremely sensite to the choice of radial cut in this range.}

\begin{figure}
\includegraphics[]{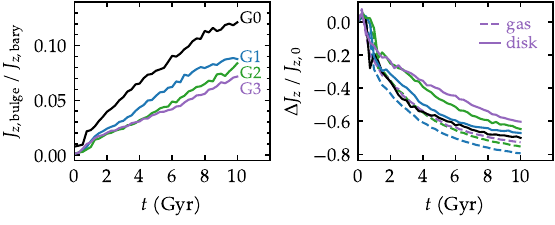}
\caption{(\textbf{a}) Fractional angular momentum held by the bulge, relative to the total angular momentum of the baryons (stellar disk, gas, and bulge), as a function of time, shown for all gas fractions of the B1 set of models. (\textbf{b}) Variation of angular momentum relative to its initial value, as a function of time, shown for all gas fractions of the B1 set of models. Solid lines represent the stellar disk; dashed lines represent the gas.}
\label{fig09}
\end{figure}

To summarize the effects of the gas on the angular momentum evolution, we take again the B1 set of models as an illustrative example.  In the left panel of Figure~\ref{fig09}, we see that the contribution of the bulge to the total angular momentum is indeed correlated to gas fraction. The gas-rich models are unable to absorb as much angular momentum as the gas-poor ones. In the right panel of Figure~\ref{fig09}, the loss of angular momentum by the disk components is similarly correlated to gas content.

It is worth emphasizing in Figure~\ref{fig09} that the decrease of angular momentum in the set of models B1 is as systematic as one would expect, given the behavior of the B1 bar strengths as a function of gas content. In other words, the loss of angular momentum is the most intense in the case of B1G0 and progressively less steep until B1G3. This applies for both the stellar disk component and also for the gas component separately. This is the fundamental reason why stronger bars develop in the models G0 and, consequently, why the rotational properties of the bulge correlate so well with gas content.

\subsection{Bulge kinematics}
\label{sec:kinematics}

In this subsection, we will quantify the rotation of the bulge in various ways and evaluate how this rotation depends on gas content.

\begin{figure}
\includegraphics[]{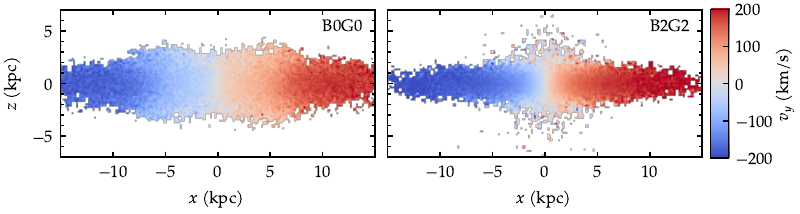}
\caption{Line-of-sight velocities of all stars (disk and bulge) in the edge-on projection. Illustrative examples are shown only for 2 galaxies at $t=10$\,Gyr.}
\label{fig10}
\end{figure}

The rotation of the stars in the galaxy as a whole is illustrated in Figure~\ref{fig10}, which displays the line-of-sight velocities when the galaxy is viewed under the edge-on projection. Here too, the bar was rotated such that its major axis is aligned with the $x$ axis. This velocity map is shown only for two examples: galaxies B0G0 and B2G2 at $t=10$\,Gyr. Galaxy B0G0 is bulgeless and has a very strong bar. Thus, when seen side-on, the box/peanut shape is quite pronounced. Galaxy B2G2 has a classical bulge, which is included in the visualization of Figure~\ref{fig10}. The bar of B2G2 is rather weak, so the morphology of the edge-on projection is dominated in the central region by the classical bulge itself. In the presence of the stellar disk, the rotation of the bulge is rather subtle to notice, as the rotational velocities in Figure~\ref{fig10} range to as much as $\sim$200\,km\,s$^{-1}$.

\begin{figure}
\includegraphics[]{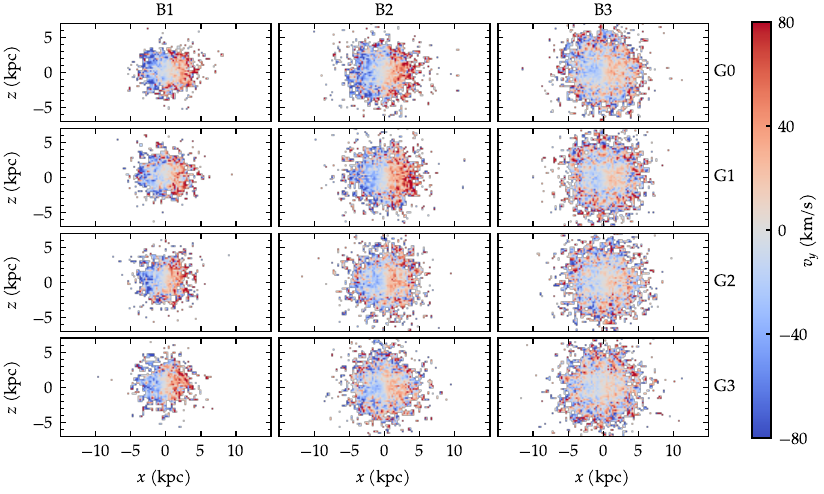}
\caption{Line-of-sight velocities of the bulge in the edge-on projection, at $t=10$\,Gyr. The columns correspond to different bulge fractions; the rows correspond to different gas fractions.}
\label{fig11}
\end{figure}

To highlight the kinematics of the bulge itself, Figure~\ref{fig11} displays velocity maps only for the bulge component, and with a suitable velocity range, at $t=10$\,Gyr. The rotation of the bulge is already quite noticeable by eye in Figure~\ref{fig11}, and the velocity range is not negligible. In many models, the bulge rotation is noticeable because it is possible to distinguish in this visualization the characteristic dipole configuration of positive/negative velocities. In the non-barred models, it is possible to note that the left and right sides of the bulge are on average indistinguishable, meaning that there is not net rotation but just isotropic velocities. Nevertheless, some small non-zero rotation can be noticed in the innermost regions. The fact that the bulges of models B3 appear to be spatially more extended than those of B1 in this visualization is not physically meaningful. It is merely a consequence of the more massive bulges being sampled by a larger number of particles, since all particles have the same mass.

\begin{figure}
\includegraphics[]{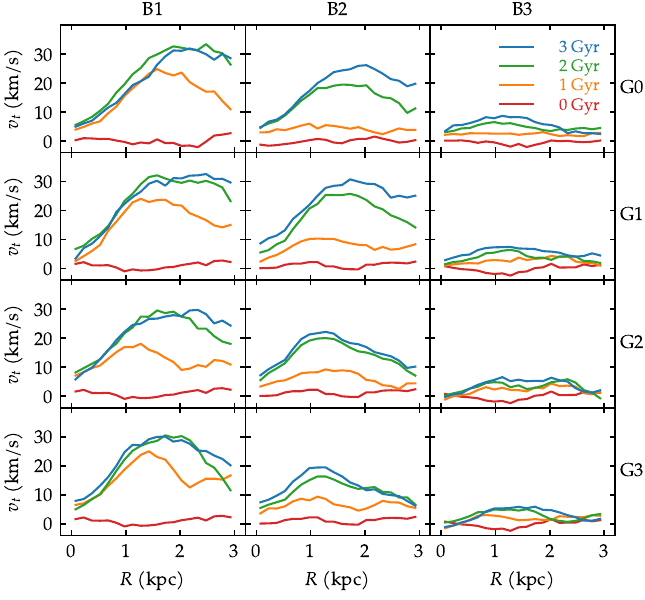}
\caption{Radial profiles of the bulge tangential velocities, shown at selected times. The columns correspond to different bulge fractions; the rows correspond to different gas fractions.}
\label{fig12}
\end{figure}

A more quantitative measurement of bulge rotation is presented in Figure~\ref{fig12}, which shows the tangential component of the velocity of bulge particles averaged within radial bins. In the beginning of the simulation, naturally there is no rotation. In the models with strong bars, a rotational velocity soon settles in. Again we find that the rotation is generally more noticeable in the galaxies with small gas content and/or low-mass bulges. {(Alternate choices of the oversampling factor in the radial bins of Figure~\ref{fig12} control the level of noise of the resulting measurements; typical choices in the range 0.5--2 translate into fluctuations smaller than 15\% around these resulting curves.)}

\begin{figure}
\includegraphics[]{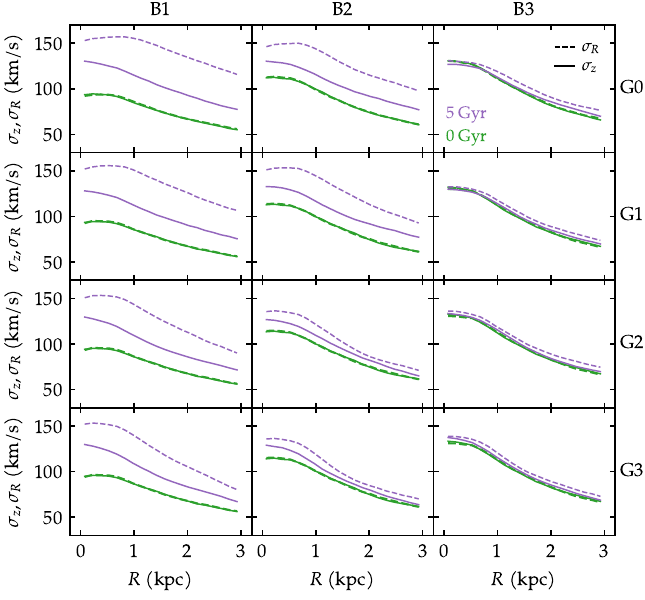}
\caption{Radial profiles of the vertical (solid lines) and radial (dashed lines) velocity dispersions of the stars in the inner region. The columns correspond to different bulge fractions; the rows correspond to different gas fractions.}
\label{fig13}
\end{figure}

\begin{figure}
\includegraphics[]{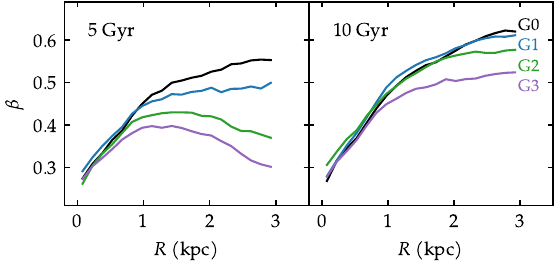}
\caption{Radial profiles of the anisotropy parameter, shown for all gas fractions of the B1 set of models at (\textbf{a}) $t=5$\,Gyr and (\textbf{b}) $t=10$\,Gyr.}
\label{fig14}
\end{figure}

Since the bulge is spherical in the initial conditions, its velocity dispersions are initially isotropic. As the bulge gains rotation, it undergoes important structural changes, including the establishment of anisotropic velocity dispersions. In Figure~\ref{fig13}, we present the velocity dispersions at a later time in the evolution ($t=5$\,Gyr), compared to the initial dispersions. At first (green lines in Figure~\ref{fig13}), the radial and vertical velocity dispersions coincide. At a later time (purple lines), there is nearly no change in the case of galaxies with weak bars. These non-rotating bulges remain isotropic. However, in the cases with strong bars, both velocity dispersions increase, with the radial dispersion (dashed lines) departing even more from its initial state than the vertical dispersion (solid lines). An anisotropy parameter may be defined as:
\begin{equation}
\beta = 1 - \left( \frac{\sigma_z}{\sigma_R} \right)^2,
\end{equation}
where $\sigma_z$ and $\sigma_R$ are the vertical and radial velocity dispersions, respectively. A $\beta$ parameter of zero corresponds to isotropy in these dimensions. In Figure~\ref{fig14}, we display the profiles of anisotropy for two moments in the simulation, $t=5$\,Gyr and $t=10$\,Gyr. This is shown for the set of models B1 to again highlight the effect of gas. We find that the settling of anisotropy depends systematically on the gas fraction as well. 

\subsection{Bulge shape}

Finally, in this subsection we investigate the shapes of the classical bulges, which begin with spherical symmetry in the initial conditions.

In order to measure the shapes of the bulges throughout the evolution, we employ the inertia tensor, which can be built easily from the masses $m_i$ and cartesian coordinates $x_i, y_i, z_i$ of each bulge particle:
\begin{equation}
\mathbf{I} = \begin{bmatrix} 
I_{xx} & I_{xy} & I_{xz} \\ 
I_{xy} & I_{yy} & I_{yz} \\ 
I_{xz} & I_{yz} & I_{zz} 
\end{bmatrix},
\end{equation}
where the diagonal elements are:
\begin{equation}
I_{xx} = \sum_i^N m_i (y_i^2 + z_i^2), \quad
I_{yy} = \sum_i^N m_i (x_i^2 + z_i^2), \quad
I_{zz} = \sum_i^N m_i (x_i^2 + y_i^2),
\end{equation}
and the off-diagonal elements are:
\begin{equation}
I_{xy} = -\sum_i^N m_i x_i y_i, \quad
I_{xz} = -\sum_i^N m_i x_i z_i, \quad
I_{yz} = -\sum_i^N m_i y_i z_i.
\end{equation}
Once the eigenvalues of $\mathbf{I}$ are obtained (in the order $\lambda_1 < \lambda_2 < \lambda_3$), then the axis ratios are computed as:
\begin{equation}
\frac{b}{a} = \sqrt\frac{\lambda_1-\lambda_2+\lambda_3}{\lambda_2 - \lambda_1 + \lambda_3}
\end{equation}
\begin{equation}
\frac{c}{a} = \sqrt\frac{\lambda_1+\lambda_2-\lambda_3}{\lambda_2 - \lambda_1 + \lambda_3}.
\end{equation}

\begin{figure}
\includegraphics[]{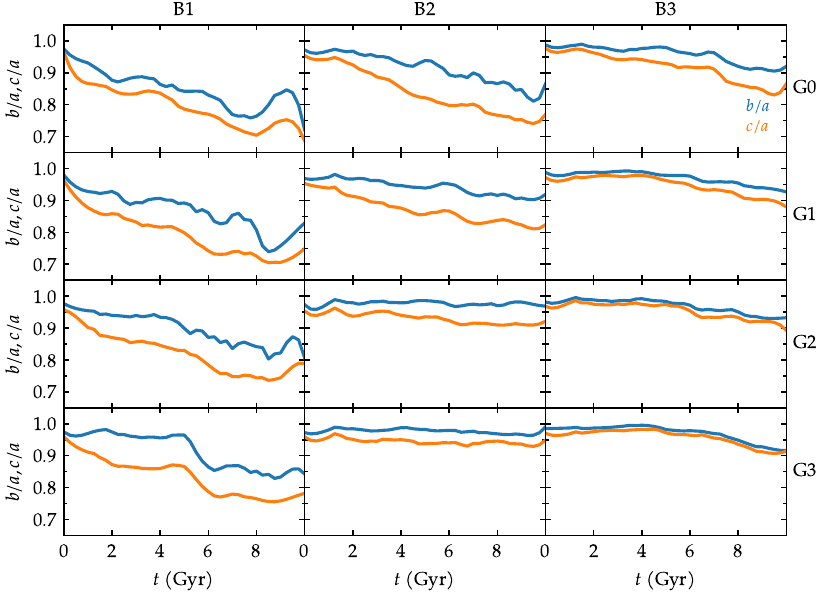}
\caption{Axis ratios of the bulge as a function of time: minor-to-major (orange) and intermediate-to-major (blue). The columns correspond to different bulge fractions; the rows correspond to different gas fractions.}
\label{fig15}
\end{figure}

{We aim to obtain the shape parameters for the bulge as whole, at each time. However, if all of the bulge particles are taken into account, a few outliers at large radii would contaminate the measurements. At the same time, applying an arbitrary spherical cut at given radius could introduce a bias towards sphericity. To avoid these two potential pitfalls, we apply the following procedure. We compute the local density at the position of each bulge particle. Then, the bulge particles are sorted in order of decreasing density. The 85\% particles of highest density are kept, and the remaining low-density bulge particles are disregarded in the shape calculations. This is approximately equivalent to performing a radial cut at roughly $R < 5$\,kpc, but with the advantage of allowing the distribution of selected particles to retain its natural ellipsoidal shape at the edges.}

The result of measuring $b/a$ and $c/a$ in this manner as a function of time is shown in Figure~\ref{fig15}. The minor-to-major axis ratios $c/a$ (orange lines) always decrease more steeply and may reach values close to $c/a\sim0.7$ in the most extreme cases. In general, all bulges tend to become slightly flattened in time, even the ones in non-barred galaxies. However, the bulges that become more flattened tend to be the ones in strongly barred galaxies. This is understandable, since those are the bulges that will have acquired the most intense rotation. Besides the vertical flattening, the intermediate-to-major axis ratio $b/a$ (blue lines) also decreases, but not as intensely. This means that the final shape of the classical bulge tends to be neither spherical nor oblate, but rather triaxial.

\begin{figure}
\includegraphics[]{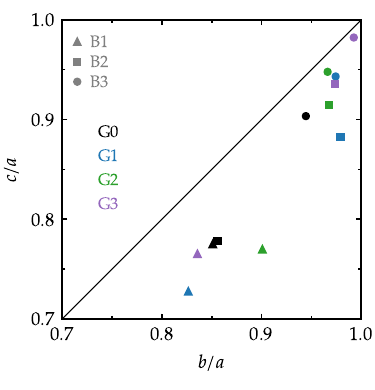}
\caption{Axis ratios of the bulge for all galaxies at $t=6$\,Gyr. The symbols represent different bulge fractions; the colors represent different gas fractions.}
\label{fig16}
\end{figure}

A diagram that helps clarify the shapes is shown in Figure~\ref{fig16}, as the $c/a$ versus $b/a$. The moment shown in this figure is $t=6$\,Gyr. A perfectly spherical bulge would be found at the upper right corner ($b/a=c/a=1$) of Figure~\ref{fig16}. Indeed, galaxy B3G3 is found closest to that configuration, because its shape was the one that deviated the least from spherical. Oblate spheroids ($a=b>c$) would lie along the rightmost edge of the figure with varying degrees of vertical flattening. Indeed, some of the B2 and B3 galaxies are not far from such regime. The diagonal line would represent a prolate spheroid ($a>b=c$). However, the resulting bulges are all in the generically triaxial regions of the diagram. It is noticeable that most of the B1 galaxies are the ones with the greatest flattenings in both dimensions. In other words, these are bulges that deviated the most from spherical symmetry.

\begin{figure}
\includegraphics[]{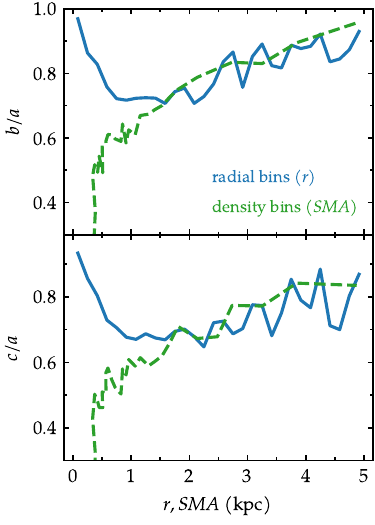}
\caption{Axis ratios of the bulge for galaxy B1G0 at $t=10$\,Gyr. The measurements are shown for equally spaced radial bins, as a function of radius (blue); and for density bins, as a function of semi-major axis (green).}
\label{fig17}
\end{figure}

{Since the bulges were found to be generally triaxial at the end of the simulations, the question arises about the adequacy of assuming cylindrical symmetry when measuring the kinematic properties in Section~\ref{sec:kinematics}. It becomes necessary to evaluate whether measuments as a function of distance from the center are appropriate when characterizing such a triaxial structure. To inspect this, we have measure the profiles of axis ratios $b/a$ and $c/a$ in two ways: with radial bins, and with density bins. In the first approach, the bulge particles are divided into equally-space radial bins, meaning that each bin is a spherical shell. The shapes are then measured for the particles inside each shell. In the second approach, the bulge particles are sorted in order of decreasing local density. Then, they are divided into density bins containing the same number of particles. The shapes are measured for the particles inside each density bin. The advantage of this approach is that a spherical geometry is nor imposed on the subsamples of particles. Rather, each density bin is allowed to retain its natural ellipsoidal shape, following the isodensity surfaces of the particle distribution.}

{The result of this comparison is shown in Figure~\ref{fig17}, where the axis ratios are shown as a function of distance from the center. In the case of the radial bins, the `radius` is simply the spherical radius of each shell. In the case of the density bins, the `radius` is taken to be the median $x$ coordinate of the particles that belong to that bin. Prior to all measuments, the galaxy had alrady been rotated such that the bar is aligned with the $x$ axis. Because of that, the median $x$ coordinate may be regarded directly as the semi-major axis of the ellipsoidal distribution.}

{We see in Figure~\ref{fig17} that the method of radial bins introduces a severe bias towards sphericity, but this discrepancy is mostly limited to the smallest radii of about $r<1.5$\,kpc, and even more notably at $r<0.5$\,kpc. Beyond $\sim$5\,kpc, the radial bins would again become unreliable due to the relatively small number of particles in that region. However, we find that in the range of 1.5--5.0\,kpc, the two methods are on average indistinct. Thus, these detailed measuments of shape suggest that the azymuthally averaged radial profiles of Section~\ref{sec:kinematics} should be sufficiently robust in the range 1.5--5.0\,kpc, which is where the relevant peaks of tangential velocity occur. Azymuthally averaged quantities will in general tend to attenuate asymmetric features. Nevertheless, given the degrees of triaxiality involved, the assumption of axisymmetry is not expected to meaningfully impact the kinematic measuments --- at least as long as physical conclusions are not being drawn from the innermost kpc itself. Finally, we note that the comparison of Figure~\ref{fig17} was limited to the strongly barred galaxy B1G0. In galaxies with milder triaxiality, this effect will be even less pronounced.}

\section{Discussion}
\label{sec:discussion}

In this paper, we have performed $N$-body hydrodynamical simulations to explore the effects that the gas exerts on the rotation of classical bulges in barred galaxies. To disentangle the effects of bulge and gas, we built a comprehensive set of models with different combinations of bulge mass fraction ranging from 0 to 30\%, and gas mass fraction also ranging from 0 to 30\%. These choices of fractions are meant to cover a physically plausible range of parameters typically observed in disk galaxies from early to late types.

In our simulations, the strength of the bar was influenced, independently, by these two factors: gas fraction and bulge fraction. The gasless bulgeless galaxy develops a very strong bar, while the gas-rich galaxies with massive bulges develop almost no bar at all. In other words, the presence of gas and the presence of a bulge tend to inhibit bar formation. Specifically, the correlation we found between bar strength and gas fraction is consistent with the bulgeless simulations of \cite{Athanassoula2013}. Additionally, the correlation we found between bar strength and bulge mass is consistent with the gasless simulations of \cite{Kataria2018}. In a sense, the results of this paper contribute to combining those two effects. 

Yet, some differences of approach should be noted. Contrary to \cite{Athanassoula2013}, our current simulations did not include star formation. This means that each of our simulations had a constant gas fraction, while in the simulations of \cite{Athanassoula2013}, the gas fractions decreased over time. {It is interesting to consider how the absence of time-varying gas fractions would impact the evolution of the simulated galaxies. Since high gas content tends to inhibit bar formation, one would expect that in simulations with constant gas fraction, this bar suppression effect would have the opportunity to act even more intensely and for longer periods of time. On the other hand, one should bear in mind that star formation is quite efficient in the early phases of galaxy evolution, meaning that the gas fraction decreases very rapidly during the first gigayears. For exemple, in the star-forming simulations of \cite{Athanassoula2013}, all galaxies drop to a gas fraction below 20\% already within the first 2\,Gyr of the simulation --- even the galaxies which had started with as much as 50, 75 or 100\% in initial gas fraction. In other words, after a steep early decline, all galaxies spend the remaining 8\,Gyr of their evolution with a slowly decreasing gras fraction, from $\sim$20\% to 5\%. It remains true that the early phase will have a lasting dynamical impact on the ultimate fate of the bars. Nevertheless, simulating galaxies with constant gas fractions during 8\,Gyr is not a unreasonable first approximation. It is aguable that the gas fractions of 10\% and 20\% would be the least unrealistic in this sense, while the 30\% case is somewhat extreme, and serves to accentuate the phenomenon being studied in a theoretical context. The impact of star formation on the correlations obtained in this paper might be sensitive to the adopted initial gas fractions.}

Another difference of approach is that \cite{Kataria2018} explored not only masses of bulges, but also several combinations of their concentrations. In the present work, we did not explore bulge concentrations systematically. On the contrary, since we adopted a fixed bulge scale length, our low-mass bulges are, by construction, a less concentrated. {It would be premature to speculate on the role of gas in simulations with different concentrations, because the results of the present work do not provide comparisons to draw such conclusions. Nevertheless, given the correlations in Section~\ref{sec:bar}, one would expect that, at each given bulge concentration, the gas fraction might act in the sense of decreasing bar strength as well. This an additional dimension in the parameter space, to be explored in the future.}

It should also be pointed out that the fate of bar formation in general may depend additionally on other parameters such as disk-to-halo mass ratio, halo central concentration \cite{Athanassoula2002, Athanassoula2003}, halo shape \cite{Machado2010, Iannuzzi2015}, halo spin \cite{Collier2018}, etc, but those were kept fixed in the present work. Thus the correlations between bar strength and gas or bulge obtained in this specific set of simulations cannot be guaranteed to hold necessarily for alternative choices of parameters.

{The influence of the additional parameters mentioned above may be briefly summarized as follows. Regarding disk-to-halo mass ratio, it is generally found that the bars will be stronger in cases where the mass in the inner region is dominated by the dark matter halo, i.e.~when the halo is massive and centrally concentrated \cite{Athanassoula2003}. The complex interplay between bars and non-spherical halos has been explored in numerous simulations \cite[e.g.][]{Berentzen2006, VillaVargas2009}. It is found that substantially triaxial halos tend to inhibit bar formations \cite{Machado2010}. Halo spin is another factor that has been thoroughly explored recently \cite[e.g.][]{Collier2018, Collier2019a, Collier2019b}. Some simulations indicate that for sufficiently high spin parameters, bars can cease their growth and even be dissolved. Each of these effects would play a role in determinig the ultimate fate of the bar in a given simulations. For example, it would be expected that a low mass, less concentrated triaxial spinning halo would produce a weaker bar than a massive concentrated spherical non-spinning halo. In this sense, all of the correlations in this paper would in principle be sensitive to combinations of these parameters. However, in order to isolate one given phenomenon of interest, it is necessary for pratical reasons to forego numerous other dimensions of the parameter space.}

The exchange of angular momentum is understood to be the fundamental mechanism that drives bar formation and evolution \cite{Athanassoula2003}. Angular momentum is transferred from the inner disk to the outer disk. Angular momentum is also absorbed by the spheroidal components, meaning the dark matter halo, and the classical bulge, if present. The dark matter halo is much more massive and thus the absolute angular momentum gained by the halo is generally the most important factor for bar formation. On the other hand, the bulge undergoes more impactful changes in its structure by absorbing angular momentum, because of its smaller mass and size. We found that the bulge always absorbs small amounts of angular momentum, {compared to the disk as a whole}. By the end of the simulation, the bulge typically holds of the order of $\sim$1\% of the total angular momentum of the baryons. {However, if only the region $R<4.2$\,kpc is considered, then the relative fraction of bulge angular momentum reaches as much as $\sim$20\% in the strongest cases. This region encompasses approximately 80\% of the bulge mass.} In our simulations, the models with strongest bars indeed experienced the most intense transfers of angular momentum. In the gas-rich cases, the rate of angular momentum lost by the disk is smaller.

{Angular momentum transfer tends to be more efficient when the spheroidal components are dynamically cold. One might wonder whether the inclusion of gas would have been responsible for increasing the inner velocity dispersions, thus hindering the ability of the bulge to receive angular momentum in gas-rich galaxies. That might be a possible contribution, although it it not immediately clear from the velocity dispersions of a given bulge fraction as a function of gas mass. Galaxies with massive bulges and/or high gas content have their bar formation partially suppressed. In turn, it was the presence of a strong bar that would have driven the rotation of bulge particles. Therefore, our results suggest that it is through this feedback that gas content indirectly prevents bulge rotation --- mainly though the suppression of strong bars.}

The kinematics of the bulge was characterized by maps of line-of-sight-velocity and by measuring its tangential velocity components. A clear correlation is found between bulge rotation and gas content. In gas-poor galaxies, the bulges are able to acquire more rotation. Furthermore, the bulge develops anisotropic velocity dispersions. Such radial and vertical dispersions are more strongly anisotropic in gas-poor galaxies. In gas-rich galaxies, the bulge acquires little rotation and thus remains nearly isotropic.

The velocity anisotropies are connected to the modifications of bulge shape throughout the evolution. The evolution of the axis ratios indicated that the bulges tend to become generally triaxial. The values obtained for the axis ratios are consistent with the results from the collisionless simulations in \cite{Athanassoula2002}. If the bulges became merely flattened vertically, they would be oblate. But the fact that the bulges shapes also display departures from circularity in the equatorial plane is an indication that this shape evolution might be driven by the stellar bar. In \cite{Saha2016} it was found that in the case of massive bulges at later times, a bulge-bar develops, i.e.~an elongation in the distribution of bulge particles that rotates together with the disk bar. {This so-called bulge-bar is to be uderstood as the elongation of the inner bulge itself being parallel to the disc bar. In the context of our results, this is manifested as the substantially elongated inner shape of the bulge when measured as a funcion of semi-major axis. In other words, within the inner 1\,kpc, the bulge is elongated with as much as $b/a \sim 0.4$ in the same direction as the disc bar.}

{As a results of the bulge being driven into an ellipsoidal or triaxial shape, this means that bar evolution proceeds in the presence of this additional lack of circularity in the equatorial plane. This may be analogous to bar evolution taking place in the context of intrinsically elliptical stellar discs, as in \cite{Machado2010}, although the lack of circularity due to the bulge is restricted to smaller radii, rather than the stellar disc globally. This phenomenon is also akin to the development of the so-called `halo bar', which is an elongation formed the inner dark matter halo \cite{ONeil2003, Athanassoula2005, Colin2006, Berentzen2006, Athanassoula2007}. Considering the effects of this dark matter structure adds yet another component to an already complex interplay of non-axisymmetric structures.}

{Observational results indicate that distinct stellar populations coexist in the bulge of the Milky Way: a low metallicity more spheroidal component, as well as metal-rich stars associated with the bar \cite[e.g.][]{Gonzalez2013, Zoccali2008, Zoccali2017, Ness2012}. If the kinematics of the spheroidal component is partially mixed with bar-like kinematics, then the disentanglement of these two populations becomes more challenging.}

\section{Conclusions}
\label{sec:conclusions}

In conclusion, we found that the presence of gas in simulated barred galaxies has a significant impact on the evolution of the classical bulge. Initially spherical and isotropic classical bulges may evolve into rotating anisotropic triaxial structures. The degree to which this transformation occurs depends systematically on the gas content of the galaxy. All of the rotational and structural properties of the bulge exhibit a consistent correlation with gas fraction, in the sense that larger gas fractions hinder the ability of the bulge to acquire rotation. Ultimately, the spin-up of the bulges is driven by bar strength, which in turn, is influenced by the presence of gas.

Future simulations including star formation would allow for a more realistic assessment of the role of gas in barred galaxies with bulges. {In particular, a future analysis of orbital frequencies would contribute to shed light on the fundamental physical mechanisms behind the phenomenon of transfer of angular momentum into the bulge.} Furthermore, more detailed studies are needed to disentangle the complex dynamical interplay between the coexisting components of classical bulge and box/peanut pseudobulges.

\vspace{6pt}

\authorcontributions{Conceptualization, RM; methodology, RM, KS, AW, GG; software, RM, KS, AW, GG; validation, RM; formal analysis, RM, KS, AW, GG; investigation, RM; resources, RM; data curation, RM, KS; writing---original draft preparation, RM; writing---review and editing, RM; visualization, RM, KS; supervision, RM; project administration, RM; funding acquisition, RM. All authors have read and agreed to the published version of the manuscript.}

\funding{RM acknowledges support from the Brazilian agency \textit{Conselho Nacional de Desenvolvimento Cient\'ifico e Tecnol\'ogico} (CNPq) through grants 406908/2018-4 and 307205/2021-5, and from \textit{Funda\c c\~ao de Apoio \`a Ci\^encia, Tecnologia e Inova\c c\~ao do Paran\'a} through grant 18.148.096-3 -- NAPI \textit{Fen\^omenos Extremos do Universo}. KS acknowledges support from CNPq. AW acknowledges support from \textit{Coordena\c c\~ao de Aperfei\c coamento de Pessoal de N\'ivel Superior - Brasil} (CAPES) -- Finance Code 001. GG acknowledges support from CNPq and \textit{Funda\c c\~ao Arauc\'aria}.}

\dataavailability{The raw data supporting the conclusions of this article will be made available upon reasonable request to the corresponding author.}

\acknowledgments{The authors acknowledge the National Laboratory for Scientific Computing (LNCC/MCTI, Brazil) for providing HPC resources of the SDumont supercomputer, which have contributed to the research results reported within this paper.}

\conflictsofinterest{The authors declare no conflicts of interest. The funders had no role in the design of the study; in the collection, analyses, or interpretation of data; in the writing of the manuscript; or in the decision to publish the results.}

\appendixtitles{yes}
\appendixstart
\appendix

\section[\appendixname~\thesection]{Numerical resolution}
\label{app:resolution}

{The simulations presented in this work do not use particles with identical masses. The masses of the stellar particles and the masses of the dark matter particles differ by factors of 3--5, which is typical in the literature \cite{Springel2018, Hopkins2018, Khoperskov2019, Bennett2022, Saha2012}. Nevertheless, in order to provide a resolution test, we have re-run simulation B3G3 with different numbers of particles in each component, such that all of the particle masses are identical, even the gas particles. This new simulation is labeled B3G3-em (equal masses). The galaxy B3G3 was chosen for this comparison, because that was the case with the largest contrast; all the other galaxies have smaller mass differences. Table~\ref{tab2} compares the numbers of particles and the mass resolutions for each component (halo, disk, bulge, gas) in B3G3 and B3G3-em.}

\begin{figure}[h]
\includegraphics[]{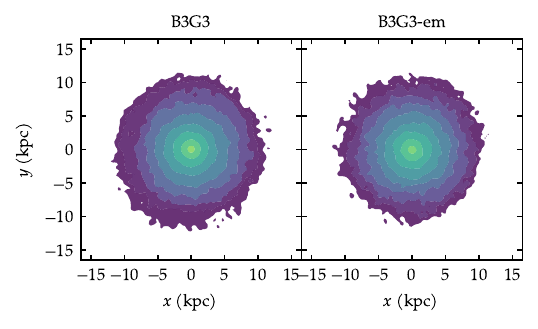}
\caption{Face-on projections of the stellar mass density, shown for galaxies B3G3 (left) and B3G3-em (right), at $t=10$\,Gyr.}
\label{fig18}
\end{figure}

\begin{figure}[h]
\includegraphics[]{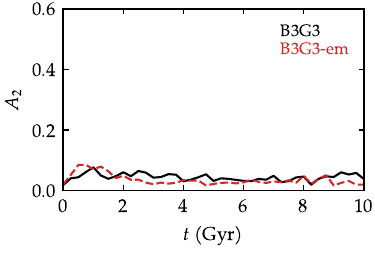}
\caption{Evolution of bar strength for galaxies B3G3 and B3G3-em.}
\label{fig19}
\end{figure}

{Figure~\ref{fig18} compares the morphology of the galaxies at the end of the simulation, which is qualitatively indistinguishable. Figure~\ref{fig19} compares the time evolution of bar strength. This quantity was chosen for this comparison, because bar strength is the most important property which controls all of the correlations found in this paper. The result is that the evolution of $A_2$ is essentially the same in both cases, within the noise that would be expected when re-running the same simulation with slightly different resolutions.}

\clearpage

\begin{table}
\caption{{Numbers of particles and mass resolutions of each component, in the comparison between the default simulation B3G3 and the version with equal masses.}}
\label{tab2}
\newcolumntype{C}{>{\centering\arraybackslash}X}
\begin{tabularx}{\textwidth}{LCCCCCCCC}
\toprule 
model & $N_{\rm h}$ & $N_{\rm d}$ & $N_{\rm b}$ & $N_{\rm g}$ & $m_{\rm h}$ & $m_{\rm d}$ & $m_{\rm b}$ & $m_{\rm g}$ \\
 &  &  &  &  & ($10^{5}\,{\rm M}_{\odot}$) & ($10^{5}\,{\rm M}_{\odot}$) & ($10^{5}\,{\rm M}_{\odot}$) & ($10^{5}\,{\rm M}_{\odot}$)  \\
\midrule
B3G3 & 1000000 & 140000 & 60000 & 200000 & 9.0 & 1.8 & 2.6 & 0.6 \\
\mbox{B3G3-em} & 3461538 & 98720 & 60000 & 42308 & 2.6 & 2.6 & 2.6 & 2.6 \\
\bottomrule
\end{tabularx}
\end{table}

\section[\appendixname~\thesection]{Resonances}
\label{app:resonances}

{We present here illustrative examples of the locations of the resonances in two galaxies. The examples are B0G0 and B2G0, in order to highlight the differences in a galaxy without bulge and one with bulge.}

{The pattern speeds of the bars, $\Omega_{\rm b}$, were measured using the method recently proposed by \cite{Dehnen2023}, which allows the measurement of $\Omega_{\rm b}$ from one single snapshot of a simulated barred galaxy. We verified, for one of the snapshots, that this method gives the same result as the direct measurements using the orientation of the bar in. The pattern speeds can only be obtained in cases where the bar is sufficiently well-defined to me realiably measured. From $t=2$\,Gyr onwards, the bars of both B0G0 and B2G0 are already strong enough to be measurable.}

\begin{figure}
\includegraphics[]{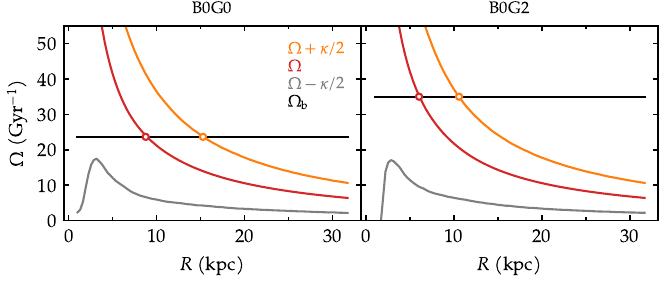}
\caption{Angular frequencies as a function of radius, used to determine the locations of the resonances. The horizontal black line corresponds to the pattern speed of the bar. The red and orange curves detemine, respectively, the corotation radius and the radius of the outer Lindblad resonance. The grey curves indicate that the inner Lindblad resonances are nonexistent in these cases. The two examples are shown for galaxies B0G0 (left) and B0G2 (right), at $t=3$\,Gyr.}
\label{fig20}
\end{figure}

\begin{figure}
\includegraphics[]{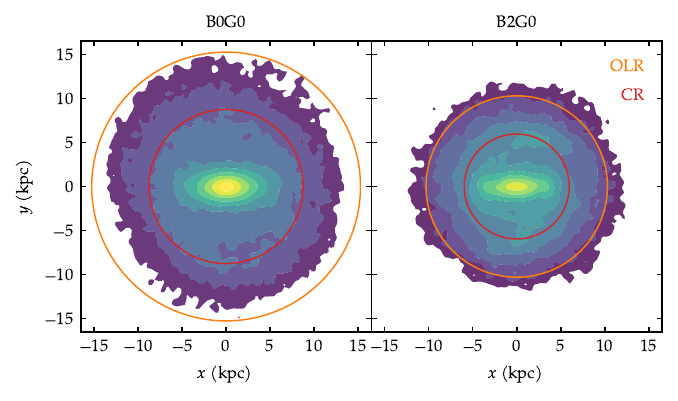}
\caption{Face-on projections of the stellar mass density, shown for two examples: galaxies B0G0 (left) and B0G2 (right), at $t=3$\,Gyr. The red and orange circles indicate, respectively, the corotation radius, and the radius of the outer Lindblad resonance, as determined from Figure~\ref{fig20}.}
\label{fig21}
\end{figure}

\begin{figure}
\includegraphics[]{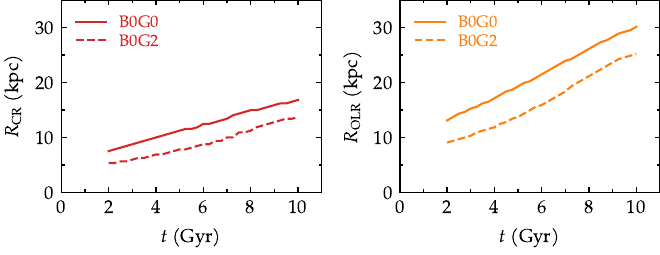}
\caption{Evolution of the corotation radius (left), and of the radius of the outer Lindblad resonance (right). The examples are shown for galaxies B0G0 (solid lines) and B0G2 (dashed lines).}
\label{fig22}
\end{figure}

{Figure~\ref{fig20} displays the $\Omega_{\rm b}$ of the two galaxies as a black horizontal line. The curves of $\Omega = v_{\rm c}/R$ were obtained by measuring the circular velocity $v_{\rm c}(R)$. The corotation radius corresponds to the radius where $\Omega_{\rm b} = \Omega_{\rm b}$. Likewise, the intercepts of $\Omega_{\rm b}$ with the curves $\Omega - \kappa/2$ and $\Omega + \kappa/2$, respectively, determine the locations of the Inner Lindblad Resonance and the Outer Lindblad Resonance. The epicycle frequency $\kappa$ at a given radius $R$ may be defined as \cite{Binney2008}:
\begin{equation}
\kappa^2 = R \frac{d \Omega^2}{dR} + 4 \Omega^2 
\end{equation}
The ILR may not exist, depending on the particular mass distribution in the center of the galaxy, which controls the shape of the circular velocity. In these cases of Figure~\ref{fig20}, the ILR is indeed nonexistent.}

{Simulations have shown that bars slow down as they become stronger \cite{Athanassoula2003}. The strong bar of B0G0 has a low $\Omega_{\rm b}$, while the relatively weaker bar of B2G0 has a larger $\Omega_{\rm b}$. Since the overall mass distributions of the two galaxies are not radically different at large radii, the consequence of this is that the CR and OLR of a strong slow bar will generally be located further out. The red and orange circles in Figure~\ref{fig21} illustrate the CR and OLR of these two example galaxies at $t=3$\,Gyr. As the bars grow stronger with time, the CR and OLR resonances will generally migrate outwards. This is what is seen in Figure~\ref{fig22}, where the time evolution of the resonances are displayed.}

{The resonances are expected to govern the angular momentum transfers that lead to the bulge acquiring rotation. However, a full spectral analysis of orbital frequencies is beyond the scope of the present study.}

\begin{adjustwidth}{-\extralength}{0cm}

\reftitle{References}
\bibliography{paper}

\PublishersNote{}
\end{adjustwidth}

\end{document}